\documentclass[twocolumn,showpacs,superscriptaddress,preprintnumbers,amsmath,amssymb]{revtex4}

\usepackage{graphicx}
\usepackage{dcolumn}
\usepackage{bm}
\usepackage{color}

\begin{document}

\newcommand{\siox}{SiO$_2$}
\newcommand{\silicate}{Si$_2$O$_3$}
\newcommand{\sicoxint}{\mbox{SiC/SiO$_2$}}
\newcommand{\rootthree}{($\sqrt{3}$$\times$$\sqrt{3}$)R30$^{\circ}$}
\newcommand{\rootthreehl}{($\sqrt{\mathbf{3}}$$\times$$\sqrt{\mathbf{3}}$)R30$^{\circ}$}
\newcommand{\sixroot}{(6$\sqrt{3}$$\times$6$\sqrt{3}$)R30$^{\circ}$}
\newcommand{\sixroothl}{(6$\sqrt{\mathbf{3}}$$\times$6$\sqrt{\mathbf{3}}$)R30$^{\circ}$}
\newcommand{\three}{\mbox{(3${\times}$3)}}
\newcommand{\four}{\mbox{(4${\times}$4)}}
\newcommand{\five}{\mbox{(5${\times}$5)}}
\newcommand{\six}{\mbox{(6${\times}$6)}}
\newcommand{\two}{\mbox{(2${\times}$2)}}
\newcommand{\twobyc}{\mbox{(2$\times$2)$_\mathrm{C}$}}
\newcommand{\twobysi}{\mbox{(2$\times$2)$_\mathrm{Si}$}}
\newcommand{\seven}{($\sqrt{7}$$\times$$\sqrt{7}$)R19.1$^{\circ}$}
\newcommand{\fourseven}{(4$\sqrt{7}$$\times$4$\sqrt{7}$)R19.1$^{\circ}$}
\newcommand{\one}{\mbox{(1${\times}$1)}}
\newcommand{\sicbar}{SiC(000$\bar{1}$)}
\newcommand{\sicbarhl}{SiC(000$\bar{\mathbf{1}}$)}
\newcommand{\bardir}{(000$\bar{1}$)}
\newcommand{\grad}{\mbox{$^{\circ}$}}
\newcommand{\cgrad}{\,$^{\circ}$C}
\newcommand{\projecta}{(11$\bar{2}$0)}
\newcommand{\projectb}{(10$\bar{1}$0}
\newcommand{\projectc}{(01$\bar{1}$0}
\newcommand{\third}{$\frac{1}{3}$}
\newcommand{\thirdspot}{($\frac{1}{3}$,$\frac{1}{3}$)}
\newcommand{\oversix}{$\frac{1}{6}$}

\newcommand{\gkdir}{$\overline{\Gamma K}$-direction}
\newcommand{\pb}{$\pi$-band}
\newcommand{\pbs}{$\pi$-bands}
\newcommand{\kv}{\mathbf{k}}
\newcommand{\kpoint}{$\overline{\textrm{K}}$ point\space}
\newcommand{\gk}{$\overline{\Gamma \mbox{K}}$}
\newcommand{\kpar}{\underline{k}$_{\parallel}$}
\newcommand{\efermi}{E$_\mathrm{F}$}
\newcommand{\edirac}{E$_\mathrm{D}$}
\newcommand{\angs}{$\mathrm{\AA}$}
\newcommand{\kk}  {\underline{k}$_{\overline{\textrm{K}}}$}
\newcommand{\kvec}{\underline{k}}

\newcommand{\red}{\textcolor[rgb]{1,0,0}}

\newcommand{\todo}[1]{\textsl{\textcolor{red}{#1}}}


\title{Large area quasi-free standing monolayer graphene on 3C-SiC(111)}

\author{C. Coletti}%
\email{c.coletti@iit.it} \affiliation{Max-Planck-Institut f\"{u}r
Festk\"{o}rperforschung, Heisenbergstr. 1, D-70569 Stuttgart,
Germany} \affiliation{Center for Nanotechnology Innovation @ NEST,
Istituto Italiano di Tecnologia, Piazza San Silvestro 12, 56127 Pisa,
Italy}

\author{K.V. Emtsev}
\affiliation{Max-Planck-Institut f\"{u}r
Festk\"{o}rperforschung, Heisenbergstr. 1, D-70569 Stuttgart, Germany}

\author{A.A. Zakharov}
\affiliation{MAX-Lab, Lund University, Box 118, Lund, S-22100, Sweden}%

\author{T. Ouisse}
\affiliation{Laboratoire des Mat\'{e}riaux et du G\'{e}nie Physique -
CNRS
UMR5628 - Grenoble INP, Minatec 3 parvis Louis N\'{e}el, BP 257, 38016 Grenoble, France}%

\author{D. Chaussende}
\affiliation{Laboratoire des Mat\'{e}riaux et du G\'{e}nie Physique -
CNRS
UMR5628 - Grenoble INP, Minatec 3 parvis Louis N\'{e}el, BP 257, 38016 Grenoble, France}%

\author{U. Starke}%
 \email{u.starke@fkf.mpg.de}
 \homepage{http://www.fkf.mpg.de/ga}
\affiliation{Max-Planck-Institut f\"{u}r Festk\"{o}rperforschung,
Heisenbergstr. 1, D-70569 Stuttgart, Germany}

\date{\today}

\begin{abstract}
Large scale, homogeneous quasi-free standing monolayer graphene is
obtained on cubic silicon carbide, i.e. the 3C-SiC(111) surface,
which represents an appealing and cost effective platform for
graphene growth. The quasi-free monolayer is produced by
intercalation of hydrogen under the interfacial,
{\sixroot}-reconstructed carbon layer. After intercalation, angle
resolved photoemission spectroscopy reveals sharp linear {\pbs}. The
decoupling of graphene from the substrate is identified by X-ray
photoemission spectroscopy and low energy electron diffraction.
Atomic force microscopy and low energy electron microscopy
demonstrate that homogeneous monolayer domains extend over areas of
hundreds of square-micrometers.
\end{abstract}


\maketitle


The unique two-dimensional electron gas properties of graphene make
it a potential candidate for future electronics. Currently, the main
techniques adopted to produce single or few-layer graphene are
mechanical exfoliation of graphite~\cite{Novoselov2004},
graphitization of silicon carbide (SiC)~\cite{Berger2004}, chemical
vapor deposition on transition metals~\cite{Sutter2008}, and chemical
synthesis~\cite{Gomez2007}. By enabling growth of large area graphene
directly on a semi-insulating substrate, thermal decomposition of SiC
is the most promising route towards a future of carbon-based
nano-electronics~\cite{Berger2004,Emtsev2009}. Hexagonal SiC
crystals, namely 4H- and 6H-SiC, provide an ideal template for
graphene growth and thus have evolved as the substrates of choice in
the past years~\cite{First2010}. In contrast, limited attention has
been given to graphene growth on cubic SiC (3C-SiC), although the
[111] orientation of this crystal would also naturally accommodate
the six-fold symmetry of graphene. With its extreme robustness and
proven biocompatibility~\cite{Coletti2007IEEE}, cubic SiC is an
appealing platform for the growth of graphene that could lead to a
new generation of microelectromechanical systems and advanced
biomedical devices. Moreover, cubic SiC can be epitaxially grown on
Si crystals and - provided the process temperatures can be
sufficiently lowered - this could reduce the production costs of
graphene.

To date, the growth of graphene on 3C-SiC(111) was attempted by
adopting the classical ultra-high-vacuum (UHV) thermal decomposition
process~\cite{Ouerghi2010APL96}. The structural and electronic
properties of such graphene were comparable to those of graphene on
SiC(0001)~\cite{Ouerghi2010PRB, Ouerghi2010APL97}. The first carbon
layer that grows on top of SiC(111) is known as zero-layer graphene
(ZLG)~\cite{Emtsev2009}. It is in part covalently bound to the SiC
substrate and hence electronically inactive. The second carbon layer
grows on top of the ZLG without interlayer bonds and acts like
monolayer graphene. Despite the good quality on the nanometer scale,
the lateral size of homogenous graphene domains produced to date on
3C-SiC(111) is limited to about 1
$\mu$m~\cite{Ouerghi2010PRB,Ouerghi2010APL97}.

In the present work we report a truly large-scale production of
epitaxial graphene on 3C-SiC(111). We demonstrate that on cubic
substrates it is indeed possible to obtain homogenous monolayer
graphene with domains extending over areas of hundreds of
square-micrometers. To achieve this result, we combine the method of
atmospheric pressure (AP) growth~\cite{Emtsev2009} with the hydrogen
intercalation technique recently developed by our
group~\cite{Riedl2009PRL}. In the first part of this work, the
morphologies of 3C-SiC(111) and of ZLG obtained using different
processing parameters are probed using atomic force microscopy (AFM).
Subsequently, ZLG is hydrogen intercalated in order to obtain quasi
free-standing monolayer graphene (QFMLG). The chemical, electronic
and structural properties of the resulting films are investigated by
means of X-ray photoemission spectroscopy (XPS), angle resolved
photoemission spectroscopy (ARPES), and low energy electron
diffraction (LEED) and microscopy (LEEM). XPS measurements were
performed using photons from a non-monochromatic Mg K$_{\alpha}$
source (h$\nu$ = 1253.6 eV). ARPES analysis was carried out using
monochromatic He II radiation (h$\nu$ = 40.8 eV) from a UV discharge
source with a display analyzer oriented for momentum scans
perpendicular to the {\gk}-direction. LEEM experiments were performed
using the ELMITEC LEEM~III instrument at beamline I311 of the MAX
radiation laboratory in Lund (Sweden).

\begin{figure}
\begin{center}
\includegraphics[width=6.0 cm]{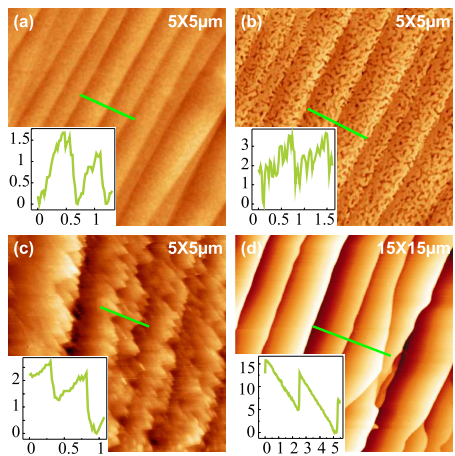}
\end{center}
\caption{(Color online) AFM micrographs and related line profile
for: (a) as-grown 3C-SiC(111), (b) UHV grown ZLG on
3C-SiC(111), (c) hydrogen etched 3C-SiC(111), (d) AP-grown
ZLG on 3C-SiC(111). Inset axes: nm (horizontal) and $\mu$m (vertical).}
\label{AFM}
\end{figure}

In order to obtain large-area homogeneous graphene, high-quality
3C-SiC(111) substrates are required. The 3C-SiC(111) samples used in
this work were grown by hetero-epitaxy on a 4H-SiC(0001) substrate
with the continuous-feed physical-vapour transport
method~\cite{Chaussende2005}. In this way, cubic epilayers with a
thickness of $\sim$390 $\mu$m were obtained and subsequently rendered
free-standing by polishing away the 4H-SiC substrate. The high
quality and crystallinity of the epilayers was verified in optical
microscopy which revealed an almost complete absence of double
positioning boundaries~\cite{Chaussende2005}. No anti-phase domain
boundaries, as those found on 3C-SiC(001)~\cite{Coletti2007APL}, were
observed. AFM analysis confirmed a high degree of order at an atomic
level. The surface consists of atomically flat terraces with 0.5 to 1
$\mu$m widths separated by steps with multiple unit cell heights (1
to 3 nm), cf. Fig.~\ref{AFM}(a). For comparison purposes, growth of
ZLG was first attempted using the UHV graphitization method described
in~\cite{Riedl2007PRB}. The resulting surface morphology is shown in
Fig.~\ref{AFM}(b). Although the original step structure remains
visible, preferential Si desorption sites make their initial
appearance, thus contributing to a rough and inhomogeneous morphology
typical of UHV prepared graphene. Notably, the zero-layer could be
removed by etching in a hydrogen atmosphere at a temperature of
1250{\cgrad} for 15 minutes. However, the terraces of the reverted
SiC surfaces present zig-zag edges with facets oriented 60{\grad}
with respect to each other, indicating etching along two preferential
directions, cf. Fig.~\ref{AFM}(c).

\begin{figure}
\begin{center}
\includegraphics[width=8.7 cm]{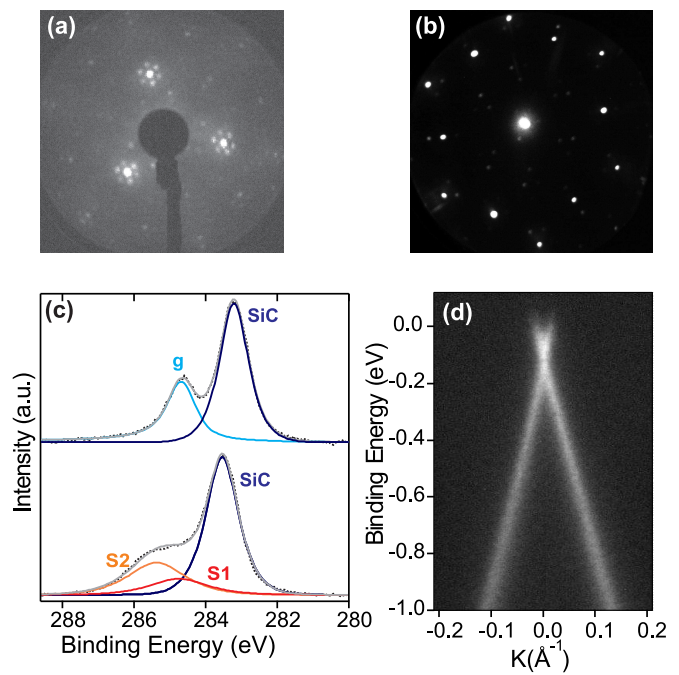}
\end{center}
\caption{(Color online) (a) LEED pattern of the {\sixroot}-reconstructed
ZLG (188 eV) (b) $\mu$-LEED image of QFMLG on 3C-SiC(111) (98 eV) (c) C1s core level emission region for
as-grown ZLG (bottom curve) and QFMLG on 3C-SiC(111) (top curve). The experimental
data are displayed in black dots. Different components, accordingly
labeled, are fitted by a line shape analysis. The gray solid line is
the envelope of the fitted components. (d) Dispersion of the {\pbs}
measured with ARPES for QFMLG
on 3C-SiC(111).} \label{XPS}
\end{figure}

The typical morphology of ZLG obtained annealing 3C-SiC(111) at
1400{\cgrad} in 800 mbar argon for 10 minutes is displayed in
Fig.~\ref{AFM}(d). The surface is highly homogeneous, with 2 to 4
$\mu$m large terraces and 6 to 15 nm high steps, which are indicative
of an increased step bunching (note the image's larger scale). The
improved morphology in panel (d) can be ascribed to the AP growth
method used~\cite{Emtsev2009}. It requires higher growth temperatures
which favor diffusion of the atoms so that the restructuring of the
surface is completed before graphene is formed. Notably, the surface
morphology shown in panel (d) could be obtained when starting from
both as-grown (panel (a)) and etched (panel (c)) surfaces. The LEED
pattern of the AP grown ZLG displayed in Fig.~\ref{XPS}(a) shows the
fractional order spots of the {\sixroot} reconstruction around the
first order diffraction spots of the SiC(111) substrate. At 188 eV
electron energy the extinction of three of these six first order
spots corroborates the complete absence of twinning domains in the
SiC substrate.

Hydrogen intercalation of the ZLG samples was achieved by annealing
at 850{\cgrad} in a molecular hydrogen atmosphere of 1
bar~\cite{Riedl2009PRL} and controlled by ARPES and XPS. Before
intercalation the ARPES analysis reveals the surface states
characteristic for the {\sixroot} reconstructed
ZLG~\cite{Riedl2009PRL,Emtsev2008} (not shown). Also the
corresponding C1s core level spectrum from XPS measurements, shown in
the bottom part of Fig.~\ref{XPS}(c), is typical for ZLG. It can be
well fitted with three components~\cite{Emtsev2008}, namely the SiC
bulk component positioned at 283.5 eV and the two zero-layer
components indicated in the literature as S1 and S2 located at 284.8
eV and 285.4 eV, respectively. The decoupling of the graphene layer
after hydrogen intercalation (weak superstructure~\cite{double}) and
its long-range order (sharp spots) are visible from the
$\mu$-LEED-pattern shown in Fig.~\ref{XPS}(b).

\begin{figure}
\begin{center}
\includegraphics[width=8.7 cm]{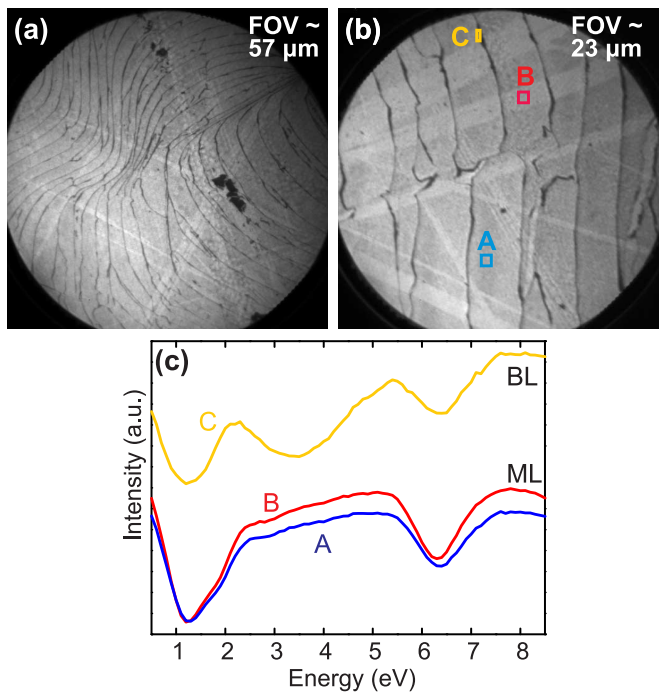}
\end{center}
\caption{(Color online)  (a) LEEM micrograph with FOV of $\sim$57 $\mu$m
recorded with an electron energy of 19 eV. (b) LEEM micrograph with
FOV of $\sim$23 $\mu$m recorded with an electron energy of 4 eV and
labeled representative region A, B and C. (c) Electron reflectivity
spectra measured for A, B and C (the curves are shifted on
the y-axis for better display).} \label{LEEM}
\end{figure}

After the intercalation process sharp monolayer {\pbs} develop as
shown in the ARPES dispersion plot in Fig.~\ref{XPS}(d). In the XPS
spectrum (see top curve of Fig.~\ref{XPS}(c)) the disappearance of
the S1 and S2 components is accompanied by a shift of the SiC
component to lower binding energies (i.e., 283.2 eV) and the
emergence of a peak located at 284.7 eV imputable to monolayer
graphene (indicated "g" in the figure). These results
indicate~\cite{Riedl2009PRL} that hydrogen atoms migrate under the
ZLG and bind to the topmost Si atoms of the 3C-SiC(111) substrate.
Hence the ZLG, freed from the covalent bonds and decoupled from the
substrate, becomes what has been named QFMLG~\cite{Riedl2009PRL}.
However, different from the case of hexagonal SiC
crystals~\cite{Riedl2009PRL}, QFMLG on 3C-SiC(111) is found to be
slightly n-doped. Both ARPES and XPS data indicate the electron
doping to be roughly 10$^{12}$ cm$^{-2}$. The origin for the doping
is currently under investigation.

Evaluation of the homogeneity of the QFMLG was carried out using
LEEM. Figure~\ref{LEEM}(a) displays a characteristic LEEM micrograph
recorded with a field of view (FOV) of $\sim$57 $\mu$m. Three
different gray levels are visible. These domains - which may be areas
of different thickness - are magnified (23 $\mu$m FOV) in panel (b),
and labeled A, B and C. For these representative regions LEEM
reflectivity spectra are plotted in panel (c). The number of minima
in such spectra in the energy regime below $\sim$6 eV identifies the
number of graphene layers~\cite{Hibino2008,Riedl2009PRL}. The spectra
indicate that the sample is homogenously covered by monolayer
graphene (regions A and B) while bilayer graphene is only present at
the step edges (region C). Note, that in contrast to pristine
epitaxial graphene, an additional minimum around 6-7 eV appears for
H-intercalated graphene~\cite{Forti_unpubl}. Hence, the lighter gray
contrast observed along certain crystal orientations (region B) is
not indicative of thickness inhomogeneities. We speculate that it
could be caused by defects in the substrate, which might be mediated
by strain. LEEM analysis also reveals that the large atomically flat
macro-terraces homogenously covered by QFMLG run uninterrupted for
hundreds of micrometers. This thickness homogeneity is highly
remarkable considering that, until now, the maximum lateral dimension
reported for homogenous graphene domains on SiC(111) was roughly 1
$\mu$m~\cite{Ouerghi2010APL97}.

In conclusion, this work demonstrates that large area QFMLG can be
produced on 3C-SiC(111) substrates. The morphological, structural and
electronic properties of such layers are fully investigated. The high
quality of the graphene obtained suggests that 3C-SiC(111) might be
an appealing and cost effective platform for the future development
of graphene technology.

C.C. acknowledges the Alexander von Humboldt Foundation for
financial support. This research was partially funded by the
European Community's Seventh Framework Programme (FP7/2007-2013)
under grant agreement no. 226716. We are indebted to the staff at
MAX-Lab (Lund, Sweden) for their advice and support.


\begin{thebibliography}{00}
\bibitem{double} The weak fractional order spots remaining can be
attributed to double diffraction between SiC and graphene.
\end{thebibliography}
\end{document}